\documentclass[%
 reprint,
 amsmath,amssymb,
 aps,
]{revtex4-2}

\usepackage{graphicx}
\usepackage{dcolumn}
\usepackage{bm}

\begin{document}

\preprint{APS/123-QED}

\title{A constraint to the uncertainty associated with the elements in the quark mixing matrix}

\author{\small Jae Jun Kim}
 
\affiliation{
 Midlands Technical College\\
 West Columbia, South Carolina\\
The United States
 \\
}%
\affiliation{
 Columbia College\\
 Columbia, South Carolina\\
 The United States\\
}%


\date{\today}

\begin{abstract}
We present an extra condition by which the size of the uncertainty associated with the diagonal elements in the quark mixing matrix can be further constrained when the matrix is parameterized by a combination of three mixing angles and a Dirac phase.  We discuss how the uncertainties of the diagonal elements can be correlated to that of the Dirac phase and what we may need to do when running fitters to constrain the uncertainties.  We discuss its implications in the end.
\end{abstract}

\maketitle


\noindent
Since the formulation of quark mixing by Cabibbo, numerous studies have been conducted to better understand the mechanisms of the mixing \cite{reference1}.  In our study, we present an extra constraint by which the size of the uncertainty associated with the elements, the diagonal elements in particular, of the mixing matrix, can be correlated to the size of a Dirac phase, taking into account the unitarity of the matrix in a standard parameterization scheme, a scheme that is based on the isospin doublet.  Then we discuss how such correlations among the size of the uncertainties can be used when addressing issues such as the tension in the first row and the column of the matrix in the quark sector.


Mathematically, we can parameterize an unitary matrix in many different ways.  Among them, there are two well-known schemes of parameterizing the matrix: Wolfenstein parameterization, which was introduced as an approximate scheme first but made to be exact later and adopted when testing the unitarity \cite{reference10}.  Another one is that illustrated in $PDG$ \cite{reference2}, where the matrix is constructed by a combination of $2 \times 2$ rotational matrices.  The two types of parameterization schemes have been widely adopted, given their suitability to measuring the size of the relevant parameters and describing the mixing.  

Understanding and imposing more stringent limits on the unitarity of the mixing matrix has been considered as one such important topic \cite{reference2}.  When trying to do so, apart from the correlation among parameters in various phase spaces, the following two conditions have been considered: First, the square sum of the elements in each column and row adds up to unity.  Second, Jarlskog invariant \cite{reference3}, needs to be held.
Imposing the two conditions along with that in different phase spaces has led us to constrain the size of the elements in the mixing matrix further when running fitters \cite{reference4}. 

Interestingly, most of the rules regarding the unitarity can be applied to the size of the central value of the elements of the matrix or that of the phase space associated with the central values.  Many studies focused on parameterizing the matrix that suits the phase space of the measurable quantities as functions of the parameters used when constructing the matrix.

That being said, there are not many studies conducted to explore conditions associated with the unitarity to that of the size of the uncertainty associated with the elements in the matrix.  Can we possibly constrain the size of uncertainty, either in their leading-order or exact?  There, our study began.

We focus on exploring a possible correlation that exists among the size of the uncertainty associated with each element, not the central value, when parameterizing it in a combination of three $\theta$s and one $\delta$, the standard parameterization in \cite{reference2}, where unitarity is guaranteed.  For that, first we describe what we can do with the uncertainty part of the matrix using the two conditions mentioned above.  After that, we study the correlation among the size of the uncertainties and its relationship to some of the issues such as the tension in the first row and column of the matrix in the quark sector.  

When we parameterize the matrix, it is known that we need at least four parameters to be used, by reducing the number to half, demanding the unitarity, along with the conditions of rephrasing the phase.  With the minimum number of parameters, the entire matrix, including the size of uncertainty associated with the matrix elements, can be constructed as a function of four independent parameters and the uncertainty associated with them.  

Note that, when we describe mixing by a combination of three $2 \times 2$ rotational matrices under the standard parameterization scheme, the permutations among the parameters need to be considered.  In other words, the order by which $2 \times 2$ matrix comes matter when constructing the whole $3 \times 3$ mixing matrix \cite{reference3}.  For instance, when we construct the matrix by three $\theta$s and one $\delta$, we need to consider at least six permutations.  As illustrated in \cite{reference3}, they are,
\begin{equation}
\begin{split}
&U = U_{1} \hspace{0.1mm} U_{2} \hspace{0.1mm} U_{3} = U_{1} \hspace{0.1mm} U_{3} \hspace{0.1mm} U_{2} = U_{2} \hspace{0.1mm} U_{1} \hspace{0.1mm} U_{3}
= U_{2} \hspace{0.1mm} U_{3} \hspace{0.1mm} U_{1}\\
&= U_{3} \hspace{0.1mm} U_{1} \hspace{0.1mm} U_{2} = U_{3} \hspace{0.1mm} U_{2} \hspace{0.1mm} U_{1}
\end{split}
\end{equation}
, where $U_{1,2,3}$ are $2 \times 2$ rotational matrices, parameterized by three $\theta$s and one $\delta$.

As a matter of fact, there are nine different ways to parameterize the matrix within the scheme of constructing the matrix with the three angles and the one phase \cite{reference6}.  It is nine since we need to consider the presence of the Dirac phase in addition to the permutation among the rotational matrices.  This happened to be not an issue in parameterization schemes such as Wolfenstein parameterization, since the parameterization is not based on rotations but based on scalings \cite{reference8}.

It is not too difficult to see that all six cases of the permutation return different sizes of three $\theta$ s and one $\delta$ but the same size of the elements in the unitary mixing matrix and that of $J$, and the uncertainties associated with the elements.  There, our question arises: Are there functions in which the size of the uncertainty, or the relative uncertainty in some combinations of the elements, can stay invariant, either in the leading order or exact, when the unitarity of the mixing matrix is assumed by the two main conditions mentioned?

Interestingly, there is one such function that we could identify, not directly the size of the uncertainty, but related to the uncertainty of the ratio of the elements, which can be utilized to constrain the size of the uncertainty among the elements further to some extent.  The one we explore in this work can begin with a combination of four sine functions.  Taking into account the expressions that represent the size of the matrix elements in \cite{reference3}, it can be written as
\begin{equation}
\begin{split}
S_i = sin \theta^{}_{23,i} sin \theta_{12,i} sin \theta_{13,i} sin \delta_{13,i}
\end{split}
\end{equation}
, where $i$ runs from 1 to 6, representing permutation in Equation 1.  Note that we only consider a Dirac phase as a component in the expression.

Writing a $3 \times 3$ unitarity matrix as,
\begin{equation}
U=
\begin{array}{c|ccccc|c}
&& V_{1} & V_{12} & V_{13} && \\
&& V_{21} & V_{2} & V_{23} && \\
&& V_{31} & V_{32} & V_{3} && \\
\end{array}
\end{equation}
, then it is not too difficult to see $S_i$ is not invariant, given that three $\theta$s and one $\delta$ can have different values when the rotational matrices are ordered differently.  

However, in all six cases, $S_i$ can be written as a function of a combination of Jarlskog invariant, along with the two out of three diagonal elements in the matrix as,
\begin{equation}
\begin{split}
&J = S_{1} \hspace{0.2mm} V_{1} \hspace{0.2mm} V_{3} = S_{2} \hspace{0.2mm} V_{1} \hspace{0.2mm} V_{2} = S_{3} \hspace{0.2mm} V_{2} \hspace{0.2mm} V_{3} = S_{4} \hspace{0.2mm} V_{1} \hspace{0.2mm} V_{2}\\
&= S_{5} \hspace{0.2mm} V_{2} \hspace{0.2mm} V_{3} = S_{6} \hspace{0.2mm} V_{1} \hspace{0.2mm} V_{3}
\end{split}
\end{equation}
, where the six cases represent the six different permutations of ordering the $2 \times 2$ matrices in Equation 2.  Note that we use a single index for the diagonal elements,
\begin{equation}
V_{1} = V_{11}, V_{2} = V_{22}, V_{3} = V_{33} 
\end{equation}
.  Given the invariance of $J$, the size of the ratio of $S$ for the six different cases of permutation can be written as a ratio of the size of the two elements in the matrix.  For instance, taking the ratio of the first three cases of permutations, $S_{1} \hspace{0.2mm} V_{1} \hspace{0.2mm} V_{3}$, $S_{2} \hspace{0.2mm} V_{1} \hspace{0.2mm} V_{2}$ and $S_{3} \hspace{0.2mm} V_{2} \hspace{0.2mm} V_{3}$, we have,
\begin{equation}
\begin{split}
A^{}_1 = V^{}_3 \hspace{0.4mm} V^{-1}_2, \hspace{0.2mm}
A^{}_2 = V^{}_1 \hspace{0.4mm} V^{-1}_2, \hspace{0.2mm}
A^{}_3 = V^{}_1 \hspace{0.4mm} V^{-1}_3
\end{split}
\end{equation}
, where $J$ and one of $V_i$ cancel out in Equation 5, when taking the ratio.  The rest of the combination of $S_i$ can also be written as a function of one of the three $A_i$.  With that, we can define $C$, the relative size of the uncertainty, the size of uncertainty with respect to that of the central value of the ratio of $S$ as,
\begin{equation}
\begin{split}
C_1 = \frac{1}{A_1} \hspace{0.2mm} \Delta A_1, \hspace{0.2mm} C_2 = \frac{1}{A_2} \hspace{0.2mm} \Delta A_2, \hspace{0.2mm} C_3 = \frac{1}{A_3} \hspace{0.2mm} \Delta A_3
\end{split} 
\end{equation}
.  Having such leads us to an interesting implication: $A_i$ is a function of $\delta_{13}$ but $V_i$ is not.  In other words, the uncertainty associated with $\delta_{13}$ depends on the size of the central value and the uncertainties of the rest of the elements.



With that being said, here comes the key point: When running the global fitters, if the size of the uncertainty of $J$ is parameterization independent, the reduction in the size of the uncertainty may need to be constrained.  As reducing the size of the uncertainties for $V_i$, the size for $\theta_{12,23,13}$ and their uncertainty can be calculated, but the size of the uncertainty for $\delta_{13}$ need to be constrained based on the rest of the parameter.








Having such patterns can be applied to some of the issues including the tension in the first row of the matrix in the quark sector.


The question of whether the matrix is an unitary or not is beyond the scope of our study.  Furthermore, running the fitters is certainly a way to constrain the size of the uncertainties of $V_i$.
However, if some of the characteristics associated with the unitarity is assumed when running the global fitter as that described in \cite{reference4} and \cite{reference9}, we may need additional constraints of their uncertainties being correlated needs to be taken into account.  The degree of impact of imposing the constraint as an extra condition may depend on how it is correlated to the constraints in different phase spaces when running the fitters.

In summary, in addition to the common constraints that can be imposed when considering a matrix to be unitary, we may need to consider additional conditions such as the correlation between the size of the uncertainty associated with the diagonal elements in the mixing matrix and the size of their reduction.  
Such could be correlated with other constraints imposed when running a global fit, so it is rather difficult to say whether it is going to alleviate the tension or make it even worse, but it is worth trying to incorporate it and see how it works when running fitters.

As for future studies, one of them could be about expanding the work to the $4 \times 4$ matrix and to a higher generation, where we set the central value for the fourth generation to zero except for the diagonal element, calculate the uncertainty size introduced due to tensions in the framework of $3 \times 3$ and see if it returns a consistent result with the theoretical estimates \cite{reference6}.




\end{document}